\documentclass[12pt,nohyper]{JHEP}

\usepackage{graphics,epsfig}

\newcommand{\Hn}{{\cal H}^{0}}
\newcommand{\Hnn}{H_{k}^{0}}

\def\ti              {\tilde}
\def\a               {\alpha}
\def\b               {\beta}
\def\d               {\delta}
\def\D               {\Delta}

\def\t               {\theta}
\def\s               {\sigma}
\def\x               {\chi}

\def\chp             {\ti \x^+}

\def\nt              {\ti \x^0}

\def\cth             {\cos\t}

\def\cst             {\cos\t_{\st}}

\def\csb             {\cos\t_{\sb}}

\newcommand{\mst}[1]   {m_{\st_{#1}}}
\newcommand{\msb}[1]   {m_{\sb_{#1}}}
\newcommand{\mstau}[1] {m_{\stau_{#1}}}
\newcommand{\mch}[1]   {m_{\ti \x^+_{#1}}}
\newcommand{\mnt}[1]   {m_{\ti \x^0_{#1}}}

\def\gev             {~{\rm GeV}}

\newcommand{\beeq}{\begin{equation}}
\newcommand{\bear}{\begin{eqnarray}}
\newcommand{\eneq}{\end{equation}}
\newcommand{\enar}{\end{eqnarray}}

\def\bvec{\left ( \begin{array}{c}}
\def\evec{\end{array} \right )}
\def\bmat{\left ( \begin{array}{cc}}
\def\emat{\end{array} \right )}

\newcommand{\gsim}{\;\raisebox{-0.9ex}
           {$\textstyle\stackrel{\textstyle >}{\sim}$}\;}

\newcommand{\lsim}{\;\raisebox{-0.9ex}{$\textstyle\stackrel{\textstyle<}
           {\sim}$}\;}

\newcommand{\msg}{m_{\tilde{g}}}

\def\sq{{\tilde{q}}} 
\def\st{{\tilde{t}}}
  
\def\sd{{\tilde{d}}}   
\def\sf{{\tilde{f}}}  
\def\stau{{\tilde{\tau}}}  
\def\snutau{{{\tilde{\nu}_\tau}}}  
\def\sb{{\tilde{b}}}
\def\sd{{\tilde{d}}}   
\def\sg{{\tilde{g}}}
\def\ch{{\tilde{\chi}}}

\preprint{
HEPHY-PUB 712/99\\
hep-ph/9903413}

\title{\vspace{2cm}
\begin{center}
Yukawa coupling corrections to\\
stop, sbottom, and stau production\\
in \boldmath{$e^+e^-$} annihilation
\end{center}}

\author{H.~Eberl, S.~Kraml, W.~Majerotto\\
Institut f\"ur Hochenergiephysik\\ 
der \"Osterreichischen Akademie der Wissenschaften,\\
A-1050 Vienna, Austria\\
\vspace{3cm}
}

\abstract{ 
We calculate within the MSSM the one--loop Yukawa coupling corrections to 
the processes
$e^+e^-\to \st_{i}\bar\st_{j},\, \sb_{i}\bar\sb_{j},\, \stau_{i}\bar\stau_{j}$ 
in order of Yukawa coupling squared. These corrections are due to the exchange
of charginos, neutralinos, charged and neutral Higgs bosons, and charged 
and neutral Higgs ghosts. We give the complete analytical formulae. We also 
perform a detailed numerical analysis of the Yukawa coupling corrections, 
including also the SUSY--QCD corrections. It turns out that for stop and sbottom 
production the Yukawa coupling corrections are typically 
$\sim \pm 10$\% of the tree--level cross section. For stau production 
they are about $\pm 5$\%. 
}

\begin{document}

\section{Introduction}
Supersymmetry (SUSY) is widely regarded as the most appealing 
extension of the Standard Model. For testing a specific model of 
supersymmetry as, for instance, the Minimal Supersymmetric Standard Model 
(MSSM) \cite{ref1,ref2,ref3} precise predictions for the production and 
decays of SUSY particles are necessary.\\
Within the last years one--loop SUSY--QCD corrections have been calculated 
for a variety of processes involving SUSY particles: for
$e^{+} e^{-} \to \sq_{i} \bar \sq_{j},\, (i,j = 1,2)$ in \cite{ref4,ref4a}, 
for $\sq_{i} \to q \ch^{0}_{k}\, (k = 1,\ldots,4)$, 
$\sq_{i} \to q' \ch^{\pm}_{l}\, (l = 1,2)$, in \cite{ref5},
for $\sq_{i} \to q \sg$ in \cite{ref6}, for $\sq_{2} \to \sq_{1} Z^{0}, 
\sq^{}_{i} \to \sq'_{j} W^{\pm}$ in \cite{ref7}, for $\sq_{2} \to \sq_{1} 
(h^{0}, H^{0}, A^{0})$, $\sq^{}_{i} \to \sq'_{j} H^{\pm}$ in \cite{ref8,ref9} 
and for the related Higgs boson decays $(h^{0}, H^{0}, A^{0}) \to \sq^{}_{i} \bar 
\sq_{j}$, $H^{\pm} \to \sq_{i}^{} \bar\sq'_{j}$ in \cite{ref8,ref10}, for
$q q' \to \sq \sq'$, $g g \to \sq \bar\sq$, $q \bar q' \to \sq \bar\sq'$, $g g 
\to \sg \sg$, $ q \bar q \to \sg \sg$, $g q \to \sg \sq$ in \cite{ref11}.
Here $\sq_{i}$ $(i=1,2)$ are the mass eigenstates of $\sq$, $\ch^{0}_{k}$ $(k = 
1,\ldots,4)$ the neutralinos, $\ch^{\pm}_{l}$ $(l = 1,2)$ the charginos, 
and $h^{0}, H^{0}, A^{0}, H^{\pm}$ the Higgs bosons of the MSSM.\\ 
The SUSY--QCD corrections have turned out to be significant, going up to 
90\%, depending on the SUSY parameters. The electroweak radiative 
corrections are expected to be much smaller as $\alpha \ll \alpha_{s}$, 
except those which originate from top, bottom and tau Yukawa couplings
\begin{equation}
	h_t = \frac{g}{\sqrt{2} m_W} \, \frac{m_t}{\sin\beta} \, , \qquad
    h_{b,\tau} = \frac{g}{\sqrt{2} m_W} \, \frac{m_{b,\tau}}{\cos\beta} \, .
 \label{yukcoup}
\end{equation}
They potentially lead 
to larger corrections due to the large top mass $m_{t}$ in $h_{t}$ and/or due 
to a large coupling $h_{b,\tau}$ in the case of large $\tan\beta$. 
Such (electroweak) Yukawa 
coupling corrections have been calculated for $e^{+} e^{-} \to \ch^{+}_{i} \ch^{-}_{j}$
in \cite{ref11a,ref11b}, for $\sb \to t \ch_{1}^{-}$ in \cite{ref12}, and for 
$H^{+} \to W^{+} A^{0}$ in \cite{ref13}. The calculations have shown that 
also these corrections can be important.\\
In this article we calculate the one--loop Yukawa coupling corrections 
to the process
 \begin{equation}
 	e^{+} e^{-} \to \sf_{i}\,\, \bar{\!\!\sf}_{\!j} \quad (i,j = 1,2), 
\quad \sf = \st, \sb, \stau \, ,
  \label{prodproc}
 \end{equation}
where $\sf_{i}$ are the mass eigenstates of $\sf$.
These corrections are due to the exchange of charginos, neutralinos, 
Higgs bosons, and Higgs ghosts. They are computed in order
$\sim h_{f}^{2}, h_{f'}^{2}, h_{f} h_{f'}$ with $f = t,b,\tau$.  
$f'$ denotes the isospin partner (e.g. $f = t , f' = b$). 
We work in the on--shell scheme. The calculation 
requires the renormalization of the sfermion mixing angle $\theta_{\sf}$,
which was first applied in \cite{ref4a}. In this paper 
we take a slightly modified renormalization condition for $\theta_{\sf}$
\cite{ref12}. In 
principle, the computation of the Yukawa coupling corrections
can be done in the unitary gauge. However, 
because of electroweak symmetry breaking the longitudinal components of 
$W^{\pm}$ and $Z^{0}$ exchange behave as the contributions from the Higgs 
ghosts $G^{\pm}$ and $G^{0}$, also being proportional to 
$h_{f}^{2}, h_{f'}^{2}, h_{f} h_{f'}$.
This is properly taken into account by using the t'Hooft--Feynman gauge 
($\xi = 1$) which we take here. 
We will see that the contributions due to the exchange of $G^{0}$ and $G^{+}$ are 
important.

\section{Tree--level formulae}

In the case of the $3^{rd}$ generation the weak eigenstates $\sf_{L}$ and $\sf_{R}$
mix to mass eigenstates $\sf_{1}$ and $\sf_{2}$ ($m_{\sf_1} < m_{\sf_2}$):
\begin{eqnarray}
\tilde{f}_1=\tilde{f}_L \cos\theta_\sf +\tilde{f}_R \sin\theta_\sf \ \ , \hspace*{1cm}
\tilde{f}_2=- \tilde{f}_L \sin\theta_\sf +\tilde{f}_R \cos\theta_\sf \, ,
\label{sftrans}
\end{eqnarray}
with the sfermion mixing angle $\theta_\sf$. 
Using the rotation matrix
\begin{equation}
	R^{\sf} = 
\left(
	\begin{array}{cc}\cos\theta_\sf & \sin\theta_\sf \\
	-\sin\theta_\sf & \cos\theta_\sf \end{array}\right)  
 \label{Rijdef}
\end{equation}
we can write eq.~(\ref{sftrans})  
in the form $\sf_i = 
R^{\sf}_{i1} \sf_L + R^{\sf}_{i2} \sf_R$.\\
The tree--level production process eq.~(\ref{prodproc}) proceeds via 
$\gamma$ and $Z$ exchange in the s--channel.
The cross section at tree--level is given by:
\begin{equation} \label{sigtree}
\sigma^0(e^+ e^- \to \sf_i \bar{\sf}_j)
=  \frac{c_\sf}{3}\,
\frac{\pi \alpha^2}{s}\,\lambda^{3/2}_{ij}\,
\left[ e_f^2 \delta_{ij} - T_{\gamma Z} e_f a_{ij}^\sf \delta_{ij}
+ T_{ZZ} {(a_{ij}^\sf)}^2 \right] \, ,
\end{equation}
with
\begin{eqnarray}
\hspace{-5mm} && T_{\gamma Z} =  \frac{v_e}
{8c_W^2s_W^2}\frac{s(s-m_Z^2)}{(s-m_Z^2)^2+ \Gamma^2_Z m_Z^2}\, ,
\quad 
T_{ZZ} = \frac{(a_e^2+v_e^2)}{256s_W^4c_W^4}
\frac{s^2}{(s-m_Z^2)^2+ \Gamma^2_Z m_Z^2} \, .
\end{eqnarray}
Here $\lambda_{ij}= (1-\mu_i^2 - \mu_j^2)^2 - 4 \mu_i^2 \mu_j^2$ with
$\mu^2_{i,j}= m^2_{\sf_{i,j}}/s$. $e_f$ is the charge of the sfermion
($e_t = 2/3, e_b = - 1/3, e_\tau = -1$). 
$c_\sf$ is a color factor, for squarks 
$c_\sf = 3$ and for sleptons $c_\sf = 1$. 
$v_e$ and $a_e$ are the vector and axial vector couplings of the electron to the
$Z$ boson: $v_e = -1 + 4 s_W^2$ 
(with $s_W \equiv \sin\theta_W$, $c_W \equiv \cos\theta_W$, $\theta_W$ is the 
Weinberg angle), $a_e = -1$, 
$a_{ij}^{\sf}$ are the relevant parts of the couplings to $Z \sf_i 
\bar{\sf}_j$, see eq.~(\ref{aijcoup}), and $\Gamma_Z$ is the total width of 
the $Z$~boson.

\section{One--loop corrections}
The corrected cross section including SUSY--QCD
corrections in ${\cal O}(\alpha_{s})$ and Yukawa couplings corrections in 
${\cal O}(h_{f}^{2},h_{f'}^{2}, h_{f} h_{f'})$ \cite{thesis} 
can be written as
\begin{equation} \label{sig1}
\sigma = \sigma^{0} + \Delta \sigma^{g} + \Delta \sigma^{\sg}
+ \Delta \sigma^{yuk}\, .
\end{equation}
The SUSY--QCD corrections $\Delta \sigma^{g}$ and $\Delta \sigma^{\sg}$ 
are given in \cite{ref4a}.
The Yukawa couplings corrections $\Delta \sigma^{yuk}$ 
can be written as a sum of contributions from exchange of
charginos, neutralinos, charged and neutral Higgs particles, and charged
and neutral Higgs ghosts, see Fig.~\ref{fig1}. 
\begin{equation}
\Delta \sigma^{yuk} = \Delta \sigma^{\ch^{+}} + \Delta \sigma^{\ch^{0}} 
+ \Delta \sigma^{H^{+}} + \Delta \sigma^{\Hn} +
\Delta \sigma^{G^{+}} + \Delta \sigma^{G^{0}}\, ,
\label{xseccont}
\end{equation}
where $\Hn$ denotes the sum of contributions from neutral Higgs bosons $h^{0}$, 
$H^{0}$, and $A^{0}$.
According to eq.~(\ref{sigtree}) the corrections can be written as:
\begin{equation} \label{dsiga} 
\Delta \sigma^{x} = \frac{c_\sf}{3}\,
\frac{\pi \alpha^2}{s}\,\lambda^{3/2}_{ij}\,
\left[ 2 e_f \Delta (e_f)_{ij}^{(x)} \delta_{ij} -
 T_{\gamma Z} (e_f \delta_{ij}\Delta a_{ij}^{(x)} +
\Delta (e_f)_{ij}^{(x)} a_{ij}^{\sf})
+ 2 T_{ZZ} a_{ij}^{\sf} \Delta a_{ij}^{(x)} \right]\, ,
\end{equation}
(no sum over $i,j$) where $x$ indicates the exchange of
$\ch^{+}, \ch^{0}, H^{+}, \Hn, G^{+}, G^{0}$.
The terms $\Delta a_{ij}^{(x)}$ and $\Delta (e_f)_{ij}^{(x)}$ are 
decomposed as:
\begin{eqnarray}
 \label{deltaasf}
\Delta a_{ij}^{(x)} &= & \delta a_{ij}^{(v,x)} + \delta a_{ij}^{(w,x)} +
\delta a_{ij}^{(\tilde\theta,x)} 
\, ,\\ \label{deltaesf}
\Delta (e_f)_{ij}^{(x)} &= & \delta (e_f)_{ij}^{(v,x)} + 
\delta (e_f)_{ij}^{(w,x)}\, .
\end{eqnarray}
The upper index $v$ denotes the vertex corrections (Fig.~\ref{fig1}a--f),
$w$ the wave--function corrections (Fig.~\ref{fig1}g--i),
and $\delta a_{ij}^{(\tilde\theta,x)}$ is the counterterm due to  
the renormalization of the mixing angle $\theta_{\sf}$.
The latter is necessary because the couplings $a_{ij}^{\sf}$ explicitly depend
on the sfermion mixing angle, see
eq.~(\ref{aijcoup}). The total correction terms $\Delta a_{ij}^{(x)}$ and
$\Delta (e_f)_{ij}^{(x)}$ are ultraviolet finite. 
In the computation we use the one--, two--, 
and three--point functions $A_{0}, B_{0}, C_{0}, C_{1}$, and $C_{2}$ 
\cite{tHooft} in the convention of \cite{Denner}.\\

\FIGURE{
	\centerline{\resizebox{14.5cm}{!}{\includegraphics{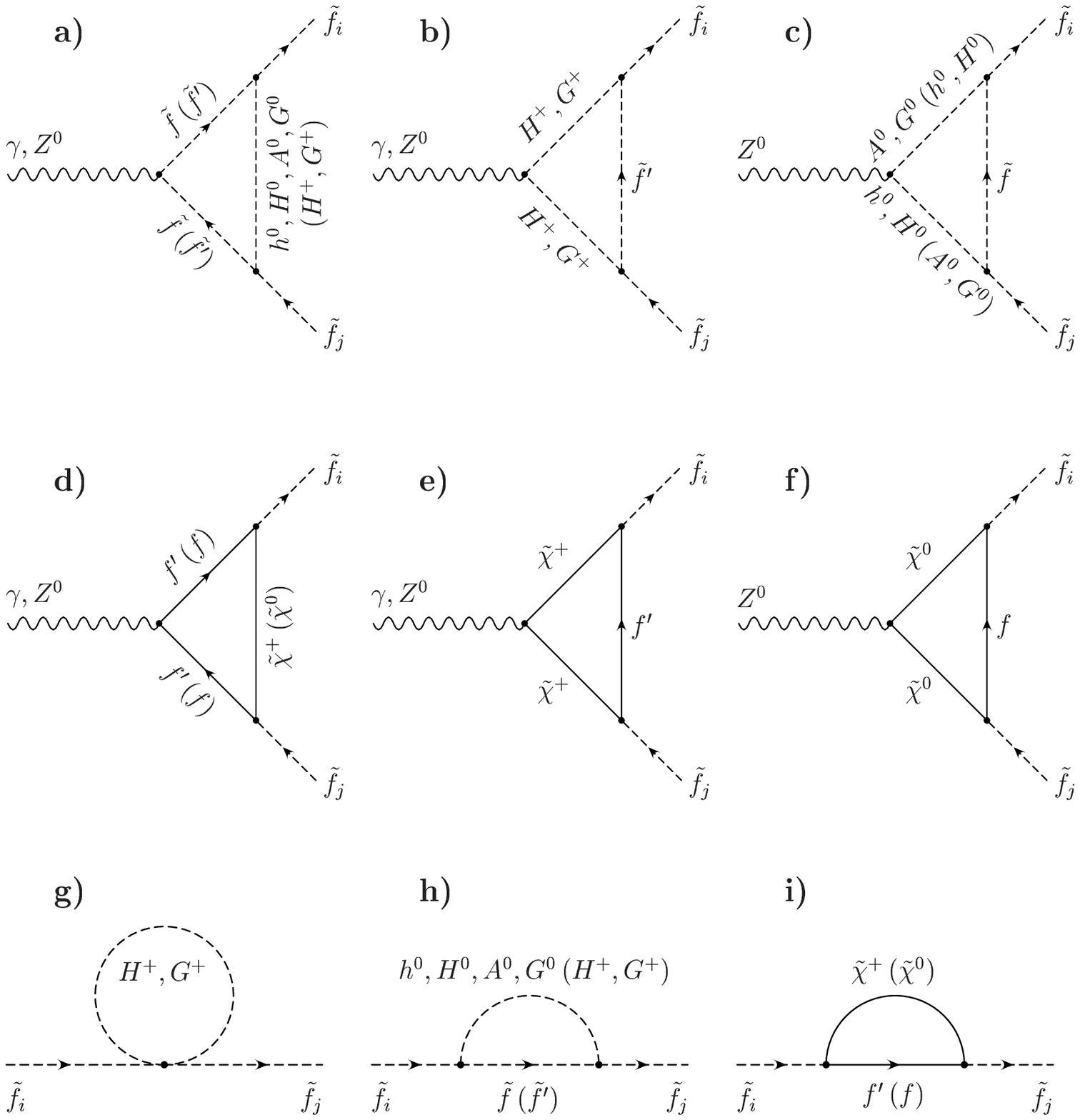}}}
    \vspace{5mm}
	\caption{Feynman diagrams for the one--loop corrections to
    $e^{+} e^{-} \to \sf_{i} \bar\sf_{j}$ in order of Yukawa couplings squared.}
	\protect\label{fig1}
}

First we calculate the
vertex corrections corresponding to Fig.~\ref{fig1}a--f. 
We introduce the variable
$A_{V}^{x}$ which is connected to the vertex correction terms
$\delta a_{ij}^{(v,x)}$, eq.~(\ref{deltaasf}), and  $\delta (e_{f})_{ij}^{(v,x)}$,
eq.~(\ref{deltaesf}), by the
relations
\begin{equation}
 \delta a_{ij}^{(v,x)} = \frac{1}{(4 \pi)^{2}}\,A_{Z}^{x}\, ,
\quad 
\delta (e_{f})_{ij}^{(v,x)} = \frac{1}{(4 \pi)^{2}}\,A_{\gamma}^{x}\, .	
  \label{AVxdef}
\end{equation} 

For the one--loop graphs of $(Z^{0*}, \gamma^{*}) \to \sf_i \bar\sf_j$ 
with  three scalars propagating in the loop, Fig.~\ref{fig1}a--c,  
i.~e. the graphs with $h^{0}$,  $H^{0}$,  $A^{0}$,
$H^{\pm}$, $G^{0}$, and $G^{\pm}$ exchange,
one has the following generic formula
\begin{equation}
	A = g_{0}\, g_{1}\, g_{2}\, (C_{0} + C_{1} + C_{2})
 \label{Ascalgener}
\end{equation}
with $(m_{\sf_i}^{2}, s, m_{\sf_j}^{2},  M_0^2, M_1^2, M_2^2)$
as the arguments of the C--functions. The couplings $g_{0}, g_{1}, g_{2}$ and
the masses $M_{0}, M_{1}, M_{2}$ of the particles in the loop can be read off from
Table~\ref{AVhiggstable}. The contributions from the ghosts $G^{+}$ and
$G^{0}$ are not given explicitly because
they can be easily calculated from the contributions with $H^{+}$ and
$A^{0}$ by using the transformation rules
\begin{equation}
 \beta \to \beta - \frac{\pi}{2}\, , \quad 
  m_{H^{+}} \to m_{W^{+}}\, , \quad 
  m_{A^{0}} \to m_{Z^{0}}\, . 	
  \label{Gtrans}
\end{equation}

\newcommand{\rb}[1]{\raisebox{1.8ex}[-1.8ex]{#1}}

\renewcommand{\arraystretch}{1.5}
\TABLE[h]{
$\begin{array}{|c||c|c|c|c|c|c|c|c|}
	\hline
		\raisebox{-1.5mm}{\mbox{name of}} & & & 
	    & \raisebox{-1mm}{$V = Z$} & 
        \raisebox{-1mm}{$V = \gamma$} &  &  & 
        \raisebox{-1.5mm}{\mbox{sum over}}  \\
	\raisebox{1.5mm}{\mbox{coeff.}} &
 \rb{$M_{0}$} & \rb{$M_{1}$} & \rb{$M_{2}$} & 
    \raisebox{1mm}{$g_0$} & 
    \raisebox{1mm}{$g_0$}  
	& \raisebox{1mm}{$g_1$} & 
      \raisebox{1mm}{$g_2$} & 
   \raisebox{1.5mm}{\mbox{indices}}\\
\hline
\hline  
A_{V}^{\Hnn} & m_{\Hnn}  & m_{\sf_{l}} & m_{\sf_{m}} & - 2 a_{lm}^{\sf} & 
-2 e_{f} \delta_{lm} & h_{f} (G_{\sf,k})_{il} & h_{f} (G_{\sf,k})_{mj} &
\begin{array}{c} k= 1,2,3 \\[-2mm]  l,m = 1,2 \end{array}\\
\hline
A_{V}^{H^{+}} & m_{H^{+}}  & m_{\sf'_{l}} & m_{\sf'_{m}} & - a_{lm}^{\sf'} &
 - e_{f'} \delta_{lm} & 
{(G_{\sf\sf'})}_{il} & {(G_{\sf\sf'})}_{jm} &  l,m = 1,2 \\
\hline
A_{V}^{A^{0} \Hnn } & m_{\sf_{l}} & m_{A^{0}} & m_{\Hnn} & - 4 \mbox{f}_{k}(\alpha\!-\! 
\beta) & 0 &  i h_{f} (G_{\sf,3})_{il} & h_{f} (G_{\sf,k})_{lj} & k,l = 
1,2 \\
\hline
A_{V}^{\Hnn A^{0}} & m_{\sf_{l}} & m_{\Hnn} & m_{A^{0}} &  4 \mbox{f}_{k}(\alpha\!-\! 
\beta) & 0 & h_{f} (G_{\sf,k})_{il} & i h_{f} (G_{\sf,3})_{lj} & k,l = 
1,2 \\
\hline
A_{V}^{H^{+} H^{-}} & m_{\sf'_{l}} & m_{H^{+}} & 
m_{H^{+}} & - 2 \cos2\theta_{W} & -1 & 
{(G_{\sf\sf'})}_{il} & {(G_{\sf\sf'})}_{jl} & l = 
1,2\\
\hline
\end{array}$
	\caption{Parameters for calculating the coefficients $A$, with 
three scalar particles in the loop, eq.~(\ref{AVxdef}). See also 
eq.~(\ref{Ascalgener}) and Fig.~\ref{fig1}a--c.
We used $\Hnn \equiv \{h^{0}, H^{0}, A^{0}\}$
and $\mbox{f}_{1} \equiv \cos$ and $\mbox{f}_{2} \equiv \sin$.}	
	\protect\label{AVhiggstable}
}
\renewcommand{\arraystretch}{1}

For the one--loop graphs of $(Z^{0*}, \gamma^{*}) \to \sf_i \bar\sf_j$ 
with a closed
fermion loop, Fig.~\ref{fig1}d--f, 
i.~e. the graphs with $\ch^{\pm}$ and $\ch^{0}$ exchange,
one has the following generic formula:
\begin{eqnarray}
A & = & \hphantom{+}\,
        M_0\, M_2\, (g_0^L g_1^L g_2^L + g_0^R g_1^R g_2^R) (C_0  + C_1 + C_2)
                       \nonumber\\
 && + \,M_0\, M_1\, (g_0^L g_1^R g_2^R + g_0^R g_1^L g_2^L) (C_0  + C_1 + C_2)
                       \nonumber\\
 && + \,M_1\, M_2\, (g_0^L g_1^R g_2^L + g_0^R g_1^L g_2^R) (C_1 + C_2)
                       \nonumber\\
 && + \,(g_0^L g_1^L g_2^R + g_0^R g_1^R g_2^L)
\Big( B_0(s, M_1^2, M_2^2) 
                       \nonumber\\
 && \hphantom{111111111}
    + M_0^2 ( 2 C_0 + C_1 + C_2) + m_{\sf_i}^{2} C_1 +  m_{\sf_j}^{2} 
C_2\Big) \, . 
 \label{Agener} 
\end{eqnarray}
Here the set of arguments for the C--functions is $(m_{\sf_i}^{2}, s, 
m_{\sf_j}^{2},  M_0^2, M_1^2, M_2^2)$. The couplings $g_{0,1,2}^{L}$ and
$g_{0,1,2}^{R}$ as well as the masses $M_{0}, M_{1}, M_{2}$ of the  
particles in the loop are given in Table~\ref{AVtable}.\\

\renewcommand{\arraystretch}{1.5}
\TABLE[h]{
	$\begin{array}{|c||c|c|c|c|c|c|c|c|c|c|c|}
	\hline
		\raisebox{-1.5mm}{\mbox{name of}} &  & & 
	    & \multicolumn{2}{c|}{V = Z} & 
        \raisebox{-1mm}{$V = \gamma$} &  &  &  &  & 
        \raisebox{-1.5mm}{\mbox{sum over}}  \\
	    \cline{5-6}
	\raisebox{1.5mm}{\mbox{coeff.}} & \rb{$M_{0}$} & \rb{$M_{1}$} & \rb{$M_{2}$} &
    g_0^R & g_0^L & 
    \raisebox{1mm}{$g_0^R = g_0^L$}  
	& \raisebox{1mm}{$g_1^R$} & 
      \raisebox{1mm}{$g_1^L$} & 
      \raisebox{1mm}{$g_2^R$} & 
  	  \raisebox{1mm}{$g_2^L$} &
   \raisebox{1.5mm}{\mbox{indices}}  \\
	\hline
	\hline
		A_{V}^{\ch^+} & m_{\ch^{+}_{l}} & m_{f'} &  m_{f'} &  
	    4 \,  C_R^{f'} & 4 \,  C_L^{f'} &  e_{f'} &
	    k_{il}^\sf  & l_{il}^\sf & l_{jl}^\sf & k_{jl}^\sf & l = 1,2   \\
	\hline
		A_{V}^{\ch^0} & m_{\ch^{0}_{l}} & m_{f} & m_{f}  &
	    4 \,  C_R^{f} & 4 \,  C_L^{f} &  e_{f} &
	   \hat b_{il}^\sf  & \hat a_{il}^\sf & \hat a_{jl}^\sf &
	   \hat b_{jl}^\sf & l = 1,\dots,4 \\
	\hline
	   A_{V}^{\ch^+\ch^-}(\st) & m_{b} & m_{\ch^{+}_{k}} & m_{\ch^{+}_{l}} &
	   - 4 \,  O^{'R}_{kl} & - 4 \,  O^{'L}_{kl} &  \delta_{kl} & 
	   k_{ik}^\st  & l_{ik}^\st & l_{jl}^\st & k_{jl}^\st & k,l = 1,2   \\
	\hline
	   A_{V}^{\ch^+\ch^-}(\sb) & m_{t} & m_{\ch^{+}_{k}} & m_{\ch^{+}_{l}} &
	    4 \,  O^{'L}_{kl} &  4 \,  O^{'R}_{kl} &  - \delta_{kl} & 
	   k_{ik}^\sb  & l_{ik}^\sb & l_{jl}^\sb & k_{jl}^\sb & k,l = 1,2   \\
	\hline
	   A_{V}^{\ch^0\ch^0} & m_{f} & m_{\ch^{0}_{k}} & m_{\ch^{0}_{l}} &
	   4 \,  O^{''L}_{kl}  & - 4 \,  O^{''L}_{kl} & 0 & 
	    \hat b_{ik}^\sf  & \hat a_{ik}^\sf & \hat a_{jl}^\sf &
	    \hat b_{jl}^\sf &  k,l = 1,\dots,4 \\
	\hline 
	\end{array}$
\caption{Parameters for calculating the coefficients $A$ with a closed 
fermion loop, eq.~(\ref{AVxdef}). See also eq.~(\ref{Agener}) and Fig.~\ref{fig1}d--f. 
The coefficient $A_{V}^{\ch^+\ch^-}(\stau)$ is obtained
from $A_{V}^{\ch^+\ch^-}(\sb)$ by using the rules $t \to \nu_{\tau}$ and
$b \to \tau$. Further we used $O^{''R} = -O^{''L}$.}	
	\protect\label{AVtable}
}
\renewcommand{\arraystretch}{1}

The sfermion wave--function renormalization due to the sfermion self--energy graphs
(Fig.~\ref{fig1}g--i)
with exchange of the particle(s) $x$ leads in the on--shell scheme to
\begin{equation}
\label{aijwave}
\delta a_{ij}^{(w,x)} =  \frac{1}{2} (\delta Z^{(x)}_{ii} + 
\delta Z_{jj}^{(x)}) a_{ij}^{\sf}
 + \delta Z^{(x)}_{i'i}  a_{i'j}^{\sf} 
+ \delta Z_{j'j}^{(x)}  a_{ij'}^{\sf} \, ,
\end{equation}
\begin{equation}
\label{efwave}
\delta a_{ij}^{(w,x)} =  \frac{1}{2} (\delta Z^{(x)}_{ii} + 
\delta Z_{jj}^{(x)}) a_{ij}^{\sf}
 + \delta Z^{(x)}_{i'i}  a_{i'j}^{\sf} 
+ \delta Z_{j'j}^{(x)}  a_{ij'}^{\sf} \, ,
\end{equation}
\begin{equation} \label{eqwave}
\delta (e_f)^{(w,x)}_{ii} = e_f \delta Z^{(x)}_{ii} \,, \quad 
\delta (e_f)^{(w,x)}_{12} = \frac{e_f}{m_{\sf_1}^2-m_{\sf_{2}}^2}
\left\{\delta Z_{12}^{(x)} +  \delta Z_{21}^{(x)} \right\} \, ,
\end{equation}
with $i \neq i'$, $j \neq j'$, 
$\delta Z^{(x)}_{ii} = 
-\mbox{Re}\left\{\Sigma_{ii}'^{(x)}(m_{\sf_i}^2)\right\}$,
where $\Sigma_{ii}'^{(x)}(m^2)= \partial \Sigma_{ii}^{(x)}(k^2)/
\partial k^2 |_{k^2=m^2}$, and
\begin{equation}
\delta Z^{(x)}_{12} = 
\frac{\Sigma_{12}^{(x)}(m_{\sf_2}^2)}{m_{\sf_1}^2-m_{\sf_{2}}^2} \, ,
\quad \mbox{and} \quad
\delta Z^{(x)}_{21} = 
- \frac{\Sigma_{12}^{(x)}(m_{\sf_1}^2)}{m_{\sf_1}^2-m_{\sf_{2}}^2} \, .	
\end{equation}
Note, that $\Sigma_{12}(k^{2}) = \Sigma_{21}(k^{2})$.\\  

The self--energy corresponding to Fig.~\ref{fig1}g is
  \begin{equation}
  \Sigma_{ij}^{(H^+)} = - \frac{1}{(4 \pi)^2}\, 
(b_f  R_{i1}^{\sf} R_{j1}^{\sf}  + 
c_f R_{i2}^{\sf} R_{j2}^{\sf})\,
 A_0(m_{H^+}^2) \, ,  	
   \label{SijH+}
  \end{equation}
with $b_{u} = c_{d} = h_{d}^{2} \sin^{2}\beta$ and
$b_{d} = c_{u} = h_{u}^{2} \cos^{2}\beta$ where $(u,d) \equiv (t,b)$ or
$(\nu_{\tau}, \tau)$. The indices $i$ and $j$ refer to the
sfermions $\sf_{i}$ and $\sf_{j}$. $R^{\sf}$ is given in
eq.~(\ref{Rijdef}).\\

Next we treat the self--energies corresponding to Fig.~\ref{fig1}h.
The results for the graphs 
with the neutral Higgs particles $h^0,\, H^0$, and $A^0$ and a sfermion in the 
loop are
  \begin{equation}
  \Sigma_{ij}^{(\Hnn \sf)}(k^2) =  \frac{2 h_f^2}{(4 \pi)^2}\,
 \sum_{l =1,2}\,
{(G_{\sf,k})}_{li}\, {(G_{\sf,k})}_{jl} B_0(k^2, m_{\Hnn}^2,m_{\sf_l}^2) \, .  	
   \label{SijHksf}
  \end{equation}
The couplings $G_{\sf,k}$ are given in eqs.~(\ref{Gijdef5}) -- 
(\ref{Gijdef10}).
Note that for $k = 1,2$ the matrices $(G_{\sf,k})$  are symmetric, therefore
${(G_{\sf,k})}_{li} = {(G_{\sf,k})}_{il}$. For the case of $A^0$ exchange the matrix
$(G_{\sf,k})$ is totally antisymmetric, 
${(G_{\sf,3})}_{li} = -{(G_{\sf,3})}_{il}$.\\
Analogously we get for the graphs with a
charged Higgs particle and a sfermion in the loop,
\begin{equation}
  \Sigma_{ij}^{(H^+\sf')}(k^2) =  \frac{1}{(4 \pi)^2}\,
 \sum_{l =1,2}\,
{(G_{\sf\sf'})}_{il}\, {(G_{\sf\sf'})}_{jl} B_0(k^2, m_{H^+}^2,m_{\sf'_l}^2) \, ,  	
   \label{SijH+sf}
\end{equation}
with the couplings $G_{\sf\sf'}$ given in eqs.~(\ref{Gijdef11}) and 
(\ref{Gijdef12}).\\
The ghost contributions $\Sigma_{ij}^{(G^+)}$, $\Sigma_{ij}^{(G^{0} \sf)}$,
and $\Sigma_{ij}^{(G^{+} \sf')}$ can be calculated from
$\Sigma_{ij}^{(H^+)}$, $\Sigma_{ij}^{(H^{0} \sf)}$,
and $\Sigma_{ij}^{(H^{+} \sf')}$, respectively, 
using eq.~(\ref{Gtrans}).\\

As to the self--energies corresponding to Fig.~\ref{fig1}i,
for chargino exchange we get
 \begin{eqnarray}
\Sigma_{ij}^{(\ch^+)}(k^2)  & = & -\frac{1}{(4 \pi)^2}\,
 \sum_{l =1,2}\, \bigg[ 2 m_{f'} m_l 
(l_{il}^\sf k_{jl}^\sf + k_{il}^\sf l_{jl}^\sf ) 
B_0(k^2, m_l^2, m_{f'}^2)  \\  
 & + & (l_{il}^\sf l_{jl}^\sf + k_{il}^\sf k_{jl}^\sf) 
\left(A_0(m_l^2) + A_0(m_{f'}^2) + (m_l^2 + m_{f'}^2 - k^2) 
B_0(k^2, m_l^2, m_{f'}^2)\right) \bigg]\, , \nonumber
  \label{Sijch} 
 \end{eqnarray}
where the index $l$ refers to the charginos $\ch^{+}_{l}$ (with $m_{l} 
\equiv m_{\ch^{+}_{l}}$).
For neutralino exchange we get
 \begin{eqnarray}
&& \hspace{-1.5mm}
 \Sigma_{ij}^{(\ch^0)}(k^2)   =  -\frac{1}{(4 \pi)^2}\,
 \sum_{k =1}^4\, \bigg[ 2 m_{f} m_k 
(\hat a_{ik}^\sf \hat b_{jk}^\sf + \hat b_{ik}^\sf \hat a_{jk}^\sf)
B_0(k^2, m_k^2, m_{f}^2)  
  \label{Sijneu}  
                   \\
  && \hspace{10mm} +\,
(\hat a_{ik}^\sf \hat a_{jk}^\sf + \hat b_{ik}^\sf \hat b_{jk}^\sf)
\left(A_0(m_k^2) + A_0(m_{f}^2) + (m_k^2 + m_{f}^2 - k^2) 
B_0(k^2, m_k^2, m_{f}^2)\right)\bigg] \, , \nonumber
 \end{eqnarray}
where the index $k$ refers to the neutralinos $\ch^{0}_{k}$ (with $m_{k} 
\equiv m_{\ch^{0}_{k}}$).\\

For the renormalization of the sfermion mixing angle $\theta_{\sf}$ we get
from eq.~(\ref{aijcoup})
\begin{equation}
\delta a_{ij}^{(\tilde\theta, x)} = \bmat 2 a_{12}^{\sf} & 
a_{22}^{\sf} - a_{11}^{\sf} \\
a_{22}^{\sf} - a_{11}^{\sf} &
- 2 a_{12}^{\sf} \emat \, \delta\theta^{(x)}_\sf\, .	
 \label{daijth}
\end{equation}
We require a process independent renormalization condition for $\theta_{\sf}$
that involves both mass eigenstates $\sf_{1}$ and $\sf_{2}$ (see also \cite{ref12}),
\begin{equation}
 \delta\theta_\sf^{(x)} = \frac{1}{2} (\delta Z_{12}^{(x)} - \delta Z_{21}^{(x)}) =
\frac{\Sigma_{12}^{(x)}(m^2_{\sf_1}) + \Sigma_{12}^{(x)}(m^2_{\sf_2})}{
2 (m^2_{\sf_1} - m^2_{\sf_2})} \, .	
\end{equation}
We want to point out that in the case of chargino exchange this antisymmetric 
combination of $\delta Z_{12}$ and $\delta Z_{21}$ is the only possible 
fixing condition for
$\delta\theta_\sf^{(\ch^{+})}$ as a function of the off--diagonal 
self--energies.

\section{Numerical results and discussion}

Let us now turn to the numerical analysis. 
As input parameters we take the MSSM parameters 
$M_{\ti Q,\ti U, \ti D,\ti L,\ti E}$, $A_{t,b,\tau}$, 
$m_A$, $\mu$, $M$, and $\tan\b$. 
$M$ is the SU(2) gaugino mass; 
for the U(1) gaugino mass $M'$ and the gluino mass 
$\msg$ we assume the GUT relations $M'=(5/3)\tan^2\t_W M$ and
$m_{\sg}=(\alpha_s(m_{\sg})/\alpha_2)M$.
Moreover, we take $m_t=175$ GeV, $m_b=5$ GeV, $m_Z^{}=91.187$ GeV, 
$\sin^2\t_W =0.232$, $m_W^{} = m_Z^{}\cos\t_W$, $\alpha(m_Z^{})=1/128.87$, 
and $\alpha_s(m_Z^{})=0.12$ 
[with $\alpha_s(Q)=12\pi/((33-2n_f)\ln(Q^2/\Lambda_{n_f}^2))$, 
$n_f$ being the number of quark flavours].
For the radiative corrections to the $h^0$ and $H^0$ masses and their 
mixing angle $\a$ ($-\frac{\pi}{2}\leq\a<\frac{\pi}{2}$ by convention) 
we use the formulae of \cite{mh0}; 
for those to $m_{H^+}$ we follow \cite{mhc} 
\footnote{Notice that \cite{mh0,mhc} have the opposite sign 
convention for the parameter $\mu$.}. 
In order to respect the experimental mass bounds from LEP2 
\cite{lep2} and Tevatron \cite{tevatron} we impose $m_{h^0} > 90\gev$, 
$\mch{1}>95\gev$, $m_{\st_1,\sb_1,\stau_1} > 85\gev$, and $\msg>200\gev$. 
Moreover, we require $\d\rho\,(\st-\sb) < 0.0012$ \cite{drhonum} 
from electroweak precision measurements 
using the one--loop formulae of \cite{Drees-Hagiwara} and 
$A_t^2 < 3\,(M_{\ti Q}^2 + M_{\ti U}^2 + m_{H_2}^2)$, 
$A_{b,\tau}^2 < 3\,(M_{\ti Q,\ti L}^2 + M_{\ti D,\ti E}^2 + m_{H_1}^2)$ 
with $m_{H_2}^2=(m_{A}^2+m_{Z}^2)\cos^2\b-\frac{1}{2}\,m_Z^2$ 
and  $m_{H_1}^2=(m_{A}^2+m_{Z}^2)\sin^2\b-\frac{1}{2}\,m_Z^2$ 
\cite{Deren-Savoy} to guarantee tree--level vacuum stability. 


For stop and sbottom production we choose 
$M_{\ti Q}=225\gev$, $M_{\ti U}=200\gev$, $M_{\ti D}=250\gev$, 
$A_t=A_b=400\gev$, $m_A=300\gev$, and $\tan\b=4$ 
as an illustrative parameter point. 
Moreover, we consider two sets of $M$ and $\mu$ values: 
a ``gaugino scenario'' with $M=200\gev$ and $\mu=1000\gev$ and 
a ``higgsino scenario'' with $M=1000\gev$ and $\mu=200\gev$.
The corresponding physical masses and mixing angles are listed 
in Tables~\ref{tab:s1} and \ref{tab:s2}. 

We first discuss the gaugino scenario of Table~\ref{tab:s1}:  
Figure~\ref{fig:one} shows the 1--loop (SUSY--QCD \cite{ref4a} and 
Yukawa coupling) corrected cross sections  
of $e^+e^-\to\st_i\bar\st_j$ and $e^+e^-\to\sb_i\bar\sb_j$ 
and the tree--level cross sections as a function of $\sqrt{s}$. 
As can be seen, the radiative corrections can have sizable effects. 
We now study the relative importance of the various contributions 
to these corrections.
In Fig.~\ref{fig:two}\,a we show the $\sqrt{s}$ dependence of the 
gluon, gluino, and Yukawa coupling corrections to $\s\,(e^+e^-\to\st_1\bar\st_1)$ 
relative to the tree--level cross section. 
The gluon correction is always positive and approaches 10\% of 
$\s^0$ for high $\sqrt{s}$. 
In contrast to that the gluino and the Yukawa coupling corrections can 
have either sign. In this example $|\D\s^{\sg}/\s^0| \lsim 4\%$;  
$|\D\s^{yuk}/\s^0|$ can go up to 6\%. 
The various contributions to the Yukawa coupling corrections are disentangled 
in Fig.~\ref{fig:two}\,b, where we plot $\D\s^x/\s^0$ with 
$x = \chp,\nt,\Hn,H^+,G$, see Eq.~(\ref{xseccont}) and 
$\Delta\sigma^{G} = \Delta\sigma^{G^{+}} + \Delta\sigma^{G^{0}}$. 
For $\sqrt{s}\lsim 830\gev$ all contributions are positive with 
those from Higgs exchange being the most important ones. 
For larger $\sqrt{s}$, $\D\s^{\Hn}$ and $\D\s^{H^+}$ 
become negative. Together, they can be up to $-60\%$ of the 
gluon correction. This comes from the large value of $\mu$ 
($\mu=1000\gev$) which directly enters the $\sq\sq H$ couplings. 
For $\sqrt{s}\sim 2$~TeV the correction due to chargino exchange 
also becomes important because $\chp_2 = \ti H^+$ with a mass of 
$\sim 1$~TeV. \\
Analogously, Figs.~\ref{fig:three}\,a and \ref{fig:three}\,b show 
$\D\s/\s^0$ and $\D\s^x/\s^0$ ($x = \chp,\nt,\Hn,H^+,G$) 
for $\st_1\bar\st_2$ + c.c. production for the gaugino scenario (Table~\ref{tab:s1}). 
In this case the Yukawa coupling correction is even more important 
than in Fig.~\ref{fig:two}. 
Together with the gluino correction it can even cancel the gluon 
correction.
The reason is that for $900\gev\lsim\sqrt{s}\lsim 2000\gev$ all relevant 
Yukawa coupling contributions are negative. 
Again there is a large correction due to neutral Higgs boson exchange.
The correction due to chargino exchange is sizable for $\sqrt{s}\sim2\mu$. 
Notice also that here the neutralino contribution plays an important r\^ole.\\
As for sbottom production, we see in Fig.~\ref{fig:four}\,a 
that Yukawa coupling corrections can be important, too, even for small $\tan\b$: 
$\D\s^{yuk}$ reaches $-17\%$ of $\s^0$. 
This is due to the top Yukawa coupling which enters the $t\,\sb\,\ch^\pm$ 
and the $\st\,\sb\, H^\pm$ couplings. Indeed, $\D\s^{H^+}$ and for 
$\sqrt{s}\sim2\mu$ also $\D\s^{\chp}$ give the main contributions to $\D\s^{yuk}$ 
as can be seen in Fig.~\ref{fig:four}\,b. 
However, for $\sqrt{s}\gsim 1700\gev$ the Yukawa coupling correction is less important 
than the gluino correction because $\D\s^{H^+}$ and $\D\s^{\chp}$ are of similar  
size but of opposite sign.

Let us now turn to the higgsino scenario of Table~\ref{tab:s2}. 
Figure~\ref{fig:five} shows the relative corrections to the cross section 
of $e^+e^-\to\st_1\bar{\st_1}$ for this scenario as a function of $\sqrt{s}$. 
Again Yukawa coupling corrections turn out to be important.  
In this case (small $\mu$), however, the dominant contributions come from 
exchanges of the lighter chargino and neutralinos. 
Higgs and ghost contributions are negligible.
Notice the spikes at $\sqrt{s}\sim 400\gev$.    
In Fig.~\ref{fig:six} we show the corrections to the cross section of 
$\sb_1\bar\sb_1$ production for the higgsino scenario. 
While the gluino correction is $\lsim 0.01\:\s^0$ the Yukawa coupling correction is 
about $0.04\:\s^0$ for $\sqrt{s}\gsim 800\gev$. 
For $\sqrt{s}\lsim 800\gev$ the contributions from $H^+$, $G^+$, and $\chp$ 
are of comparable size; for larger $\sqrt{s}$ the chargino contribution 
clearly dominates.

We finally discuss the Yukawa coupling correction to the stau production cross section. 
This correction may be important for large $\tan\b$ (large $\tau$ Yukawa coupling). 
It turns out that in this case $\D\s^{yuk}(e^+e^-\to\stau_i\bar\stau_j)$ 
is typically up to $\pm 5\%$ of the tree--level cross section. 
As an example, we plot in Fig.~\ref{fig:stau} the $\sqrt{s}$ dependence of 
the relevant Yukawa coupling correction contributions and the total correction of 
$\s(e^+e^-\to\stau_i\bar\stau_i)$, $i=1,2$, for 
$M_{\ti L}=280\gev$, $M_{\ti E}=250\gev$, $A_\tau=100\gev$, 
$M=200\gev$, $\mu=1000\gev$, $\tan\b=30$, and $m_A=300\gev$. 
This leads to $\mstau{1}=137\gev$, $\mstau{2}=355\gev$, and 
$\cth_{\stau}=0.65$. 
For the squark parameters, which are needed for the radiative 
corrections to the Higgs masses, we have taken 
$M_{\ti Q}=M_{\ti U}=M_{\ti D}=500\gev$, and $A_t=A_b=300\gev$.
In both Figs.~\ref{fig:stau}\,a and \ref{fig:stau}\,b 
Higgs boson and ghost exchanges yield the most important 
contributions because their couplings to staus directly involve 
the parameter $\mu$.
For $M=1000\gev$, $\mu=200\gev$, and the other parameters as above, 
we get $\mstau{1}=243\gev$, $\mstau{2}=293\gev$, and 
$\cth_{\stau}=0.64$. 
The Yukawa coupling correction again changes the tree--level cross section 
by $\lsim\pm 5\%$ with the dominant contributions coming from chargino 
and neutralino loops. Higgs and ghost contributions are of minor 
importance in this case.

\section{Conclusions}

We have calculated the supersymmetric Yukawa coupling corrections to 
stop, sbottom, and stau production in $e^+e^-$ annihilation. 
We have evaluated these corrections numerically for two scenarios,  
a gaugino scenario ($M\ll|\mu|$) and a higgsino scenario ($|\mu|\ll M$).  
It turns out that for stop and sbottom production the Yukawa coupling 
correction is typically $\pm 5\%$ to $\pm 10\%$ of the tree--level cross 
section. It can thus be as large as the SUSY--QCD correction. 
In the case of stau production the Yukawa coupling correction can 
also change the tree--level cross section by ${\cal O}(\pm 5\%)$. 
In the case $M\sim|\mu|$, where the individual charginos and neutralinos 
have both sizable gaugino and higgsino components, 
it may be that in addition to the graphs considered, also box graphs contribute. 
Because of the complexity of their computation, their influence will be studied 
in a separate article.\\  
In conclusion, we have shown that the corrections due to the Yukawa couplings 
are relevant for precision measurements at a future $e^+e^-$ Linear Collider. 

\acknowledgments{
We thank M. Diaz for valuable contributions in the first stage of this work.
We also thank W. Porod and A.~Bartl for discussions. 
This work was supported by the ``Fonds zur F\"orderung der 
wissenschaftlichen Forschung'' of Austria, project no. P13139--PHY.}

\begin{appendix}
\section{Couplings}
In this section we give the couplings which are necessary 
for calculating the matrix elements corresponding to the graphs
of Fig.~\ref{fig1}. The $Z^{0} \sf_{i}^{*} \sf_{j}$ coupling is proportional 
to $g/(4 \cos\theta_{W})\, a_{ij}^\sf$ with
\begin{equation}
a_{ij}^\sf = \bmat 4 ( I_f^{3L} \cos^2\theta_\sf -  e_f s_W^2) &
-2 I_f^{3L} \sin 2\theta_\sf\\
-2 I_f^{3L} \sin 2\theta_\sf &
 4 ( I_f^{3L} \sin^2\theta_\sf -  e_f s_W^2) \emat \, . 
 \label{aijcoup}
\end{equation}
With the abbreviation 
\begin{equation}
	C_{L,R}^f = I_f^{3L,R} - s_W^2 e_f \, .
 \label{cLRdef}
\end{equation}
$a_{ij}^\sf$ can also be written as
\begin{equation}
	\frac{a_{ij}^\sf}{4} = R^\sf \bmat C_L^f & 0 \\ 0 & C_R^f \emat 
(R^\sf)^T = R^\sf  \bmat I^{3L}_f & 0 \\ 0 & 0 \emat (R^\sf)^T  -
e_f s_W^2 \bmat 1 & 0 \\ 0 & 1 \emat \, .
\end{equation}
Here $I_f^{3L} = 1/2\, (-1/2)$ for up--type (down--type) fermions $f$ and
$I_f^{3R} = 0$ for all fermions. The matrix $R^\sf$ is given in eq.~(\ref{Rijdef}).\\

The Higgs--sfermion--sfermion couplings are given in 
{\cal  O}($h_t, h_b, h_\tau$). 
The Yukawa couplings $h_t, h_b$, and  $h_\tau$ are already given in 
eq.~(\ref{yukcoup}). 
The couplings for the $\Hn \sf_{i}^{*} \sf_{j}$ interactions ($\Hn \equiv 
\{h^{0}, H^{0}, A^{0} \}$, $i,j = 1,2$) are $-i \sqrt{2} h_{f} {(G_{\sf,k})}_{ij}$
with  
\begin{eqnarray}
 G_{\st,1}   & = &
R^\st . \bmat m_t \cos\alpha & \frac{1}{2} (A_t \cos\alpha + 
\mu \sin\alpha) \label{Gijdef5}\\ 
\frac{1}{2} (A_t \cos\alpha + \mu \sin\alpha) &
m_t \cos\alpha  \emat . (R^\st)^T \, ,\\[2mm]
 G_{\sd,1}   & = &
-R^\sd . \bmat m_d \sin\alpha & \frac{1}{2} (A_d \sin\alpha + 
\mu \cos\alpha)\\
\frac{1}{2} (A_d \sin\alpha + \mu \cos\alpha) &
m_d \sin\alpha  \emat . (R^\sd)^T \, ,\\[2mm]
 G_{\st,2}   & = &
 R^\st . \bmat m_t \sin\alpha & \frac{1}{2} (A_t \sin\alpha - 
\mu \cos\alpha)\\
\frac{1}{2} (A_t \sin\alpha - \mu \cos\alpha) &
m_t \sin\alpha  \emat . (R^\st)^T \, ,\\[2mm]
 G_{\sd,2}    & = &
 R^\sd . \bmat m_d \cos\alpha & \frac{1}{2} (A_d \cos\alpha - 
\mu \sin\alpha)\\
\frac{1}{2} (A_d \cos\alpha - \mu \sin\alpha) &
m_d \cos\alpha  \emat . (R^\sd)^T \, ,\\[2mm]
 G_{\st,3}    & = &
 \bmat 0 & -\frac{i}{2} (A_t \cos\beta + \mu \sin\beta)\\
\frac{i}{2} (A_t \cos\beta + \mu \sin\beta) & 0 \emat  \, ,\\[2mm]
 G_{\sd,3}   & = &
\bmat 0 & -\frac{i}{2} (A_d \sin\beta + \mu \cos\beta)\\
\frac{i}{2} (A_d \sin\beta + \mu \cos\beta) & 0 \emat 
\label{Gijdef10} \, .
\end{eqnarray}
Here $\sd$ stands for $\sb$ or $\stau$. 
The couplings for the $H^{\pm} \sf_{i}^{*} \sf'_{j}$ 
interactions are $i (G_{\sf\sf'})_{ij}$ with 
\begin{equation}
 G_{\st\sb} = 
R^\st . \bmat h_b m_b \sin\beta + h_t m_t \cos\beta & \hspace{5mm}  
h_b ( A_b \sin\beta + \mu \cos\beta)  \\
h_t ( A_t \cos\beta + \mu \sin\beta)  & \hspace{5mm} 
\frac{1}{2} \left( h_t \frac{m_b}{\cos\beta} +
 h_b \frac{m_t}{\sin\beta} \right)   \emat . (R^\sb)^T 
\label{Gijdef11} \, ,
\end{equation}
and
\begin{equation}
G_{\stau\snutau} =
 h_\tau\, \bmat
m_\tau \sin\beta \cos\theta_\stau + (A_\tau \sin\beta + \mu 
\cos\beta) \sin\theta_\stau & \hspace{5mm} 0\\
-m_\tau \sin\beta \sin\theta_\stau + (A_\tau \sin\beta + \mu 
\cos\beta) \cos\theta_\stau & \hspace{5mm} 0
\emat  \, .
\label{Gijdef12}
\end{equation}
Note that $(G_{\sf\sf'})_{ij} = (G_{\sf'\sf})_{ji}$.\\
For the interaction of $Z^{0}$ with charginos we need
\begin{eqnarray}
O_{ij}^{'L} & = & - V_{i1} V_{j1} - \frac{1}{2} V_{i2} V_{j2} + \delta_{ij} 
s_W^2 \, ,
                       \nonumber\\
O_{ij}^{'R} & = & - U_{i1} U_{j1} - \frac{1}{2} U_{i2} U_{j2} + \delta_{ij} 
s_W^2 \, ,
\label{OpLRij}
\end{eqnarray}
with the real rotation matrices $U$ and $V$ which diagonalize the chargino 
mass matrix \cite{ref2,charginos}, ($i,j = 1,2$). 
For the interaction of $Z^{0}$ with neutralinos we need
\begin{equation} 
\hspace{-0mm}
O_{kl}^{''L}  =  - O_{kl}^{''R}
 =  \frac{1}{2} \left( (-N_{k3} N_{l3}\! +\! N_{k4} N_{l4}) \cos2\beta 
-  (N_{k3} N_{l4}\! +\! N_{k4} N_{l3}) \sin2\beta \right)\, ,
\end{equation}
with the real rotation matrix $N$ which diagonalizes the neutralino mass 
matrix \cite{neutralinos}, ($k,l = 1,\ldots,4$).
One has: $\tilde\chi_k^0 = N_{kl}\, \tilde\psi_{Nl}^0$, where
$\tilde\psi_{Nl}^0 = (-i \tilde\lambda_{\gamma}, -i \tilde\lambda_{Z},
\tilde\psi^0_{H_a}, \tilde\psi^0_{H_b})$.\\

The chargino/neutralino--sfermion--fermion couplings are given in 
${\cal  O}(h_t, h_b, h_\tau)$. 
The coupling matrices $l_{ij}^\sf$ and $k_{ij}^\sf$ are 
 \begin{eqnarray}
l_{ij}^\st  = h_t V_{j2} R_{i2}^\st \, , 
& \qquad &
l_{ij}^\sb  = h_b U_{j2} R_{i2}^\sb \, ,
                        \nonumber\\
k_{ij}^\st  = h_b U_{j2} R_{i1}^\st \, , 
& \qquad & 
k_{ij}^\sb  = h_t V_{j2} R_{i1}^\sb \, ,
  \label{lkcoup}
 \end{eqnarray}
($i,j = 1,2$) and as $h_{\nu_\tau} = 0$
 \begin{eqnarray}
l_{ij}^\snutau  =  0\, , \hphantom{U_{j2} R_{i1}^\stau 1} 
& \qquad &
l_{ij}^\stau  = h_\tau U_{j2} R_{i2}^\stau \, ,
                        \nonumber\\
k_{ij}^\snutau  = h_\tau U_{j2} R_{i1}^\stau \, , 
& \qquad & 
k_{ij}^\stau  =  0 \, .
\end{eqnarray}
The couplings $\hat a_{ik}^\sf$ and $\hat b_{ik}^\sf$ are ($i,j = 1,2$)
\begin{equation}
\hat a_{ik}^\sf = h_f R_{i2}^\sf N^f_k \, , \qquad
\hat b_{ik}^\sf = h_f R_{i1}^\sf N^f_k \, .	
 \label{abcoup}
\end{equation}
$k = 1,\ldots,4$ refers to $\ch^{0}_{k}$,
$N^f_k$ is different for up--type and down--type fermions,
 \begin{eqnarray}
N^t_k  & = &  N_{k3} \sin\beta - N_{k4} \cos\beta\, ,
                        \nonumber\\
N^{b,\tau}_k  &= &  - N_{k3} \cos\beta - N_{k4} \sin\beta\, .
  \label{Nkfdef}
\end{eqnarray}

\end{appendix}



\clearpage
\renewcommand{\arraystretch}{1.4}

\vspace*{2cm}

\TABLE[h]{ 
\begin{tabular}{|llllllll|}
\hline
  $M$&$\!\!\!=\:200$, & $\mu$&$\!\!\!=\:1000$, 
                    & $\tan\b$&$\!\!\!=\:4$, & $m_A$&$\!\!\!=\:300$, \\
  $M_{\ti Q}$&$\!\!\!=\:225$, & $M_{\ti U}$&$\!\!\!=\:200$, 
                     & $M_{\ti D}$&$\!\!\!=\:250$, & $A_{t,b}$&$\!\!\!=\:400$. \\
\hline\hline
    $\mst{1}$&$\!\!\!=\:218$, 
  & $\mst{2}$&$\!\!\!=\:317$,  
  & $\cst$&$\!\!\!=\:-0.64$, & & \\
    $\msb{1}$&$\!\!\!=\:200$, 
  & $\msb{2}$&$\!\!\!=\:278$,  
  & $\csb$&$\!\!\!=\:\hphantom{-}0.79$, & & \\
    $\mch{1}$&$\!\!\!=\:196$, 
  & $\mch{2}$&$\!\!\!=\:1007$, 
  & $\msg$&$\!\!\!=\:559$, & & \\
    $\mnt{1}$&$\!\!\!=\:100$, 
  & $\mnt{2}$&$\!\!\!=\:196$,  
  & $\mnt{3}$&$\!\!\!=\:1002$, 
  & $\mnt{4}$&$\!\!\!=\:1007$, \\
    $m_{h^0}$&$\!\!\!=\:\hphantom{1}93$, 
  & $m_{H^0}$&$\!\!\!=\:305$, 
  & $m_{H^+}$&$\!\!\!=\:298$, 
  & $\sin\a$&$\!\!\!=\:-0.32$. \\
\hline
\end{tabular}
\caption{Gaugino scenario, all masses in [GeV].}
\protect\label{tab:s1}
}

\vspace*{2cm}

\TABLE[h]{ 
\begin{tabular}{|llllllll|}
\hline
  $M$&$\!\!\!=\:1000$, & $\mu$&$\!\!\!=\:200$, 
                    & $\tan\b$&$\!\!\!=\:4$, & $m_A$&$\!\!\!=\:300$, \\
  $M_{\ti Q}$&$\!\!\!=\:225$, & $M_{\ti U}$&$\!\!\!=\:200$, 
                     & $M_{\ti D}$&$\!\!\!=\:250$, & $A_{t,b}$&$\!\!\!=\:400$. \\
\hline\hline
    $\mst{1}$&$\!\!\!=\:113$, 
  & $\mst{2}$&$\!\!\!=\:368$,  
  & $\cst$&$\!\!\!=\:-0.68$, & & \\
    $\msb{1}$&$\!\!\!=\:231$, 
  & $\msb{2}$&$\!\!\!=\:252$,  
  & $\csb$&$\!\!\!=\:\hphantom{-}0.98$, & & \\
    $\mch{1}$&$\!\!\!=\:196$, 
  & $\mch{2}$&$\!\!\!=\:1007$, 
  & $\msg$&$\!\!\!=\:2393$, & & \\
    $\mnt{1}$&$\!\!\!=\:190$, 
  & $\mnt{2}$&$\!\!\!=\:202$,  
  & $\mnt{3}$&$\!\!\!=\:509$, 
  & $\mnt{4}$&$\!\!\!=\:1007$, \\
    $m_{h^0}$&$\!\!\!=\:106$, 
  & $m_{H^0}$&$\!\!\!=\:305$, 
  & $m_{H^+}$&$\!\!\!=\:309$, 
  & $\sin\a$&$\!\!\!=\:-0.31$. \\
\hline
\end{tabular}
\caption{Higgsino scenario, all masses in [GeV].}
\protect\label{tab:s2}
}

\vfill

\renewcommand{\arraystretch}{1}

\clearpage

\setlength{\unitlength}{1mm}


\FIGURE{
	\centerline{\resizebox{13.5cm}{!}{\includegraphics{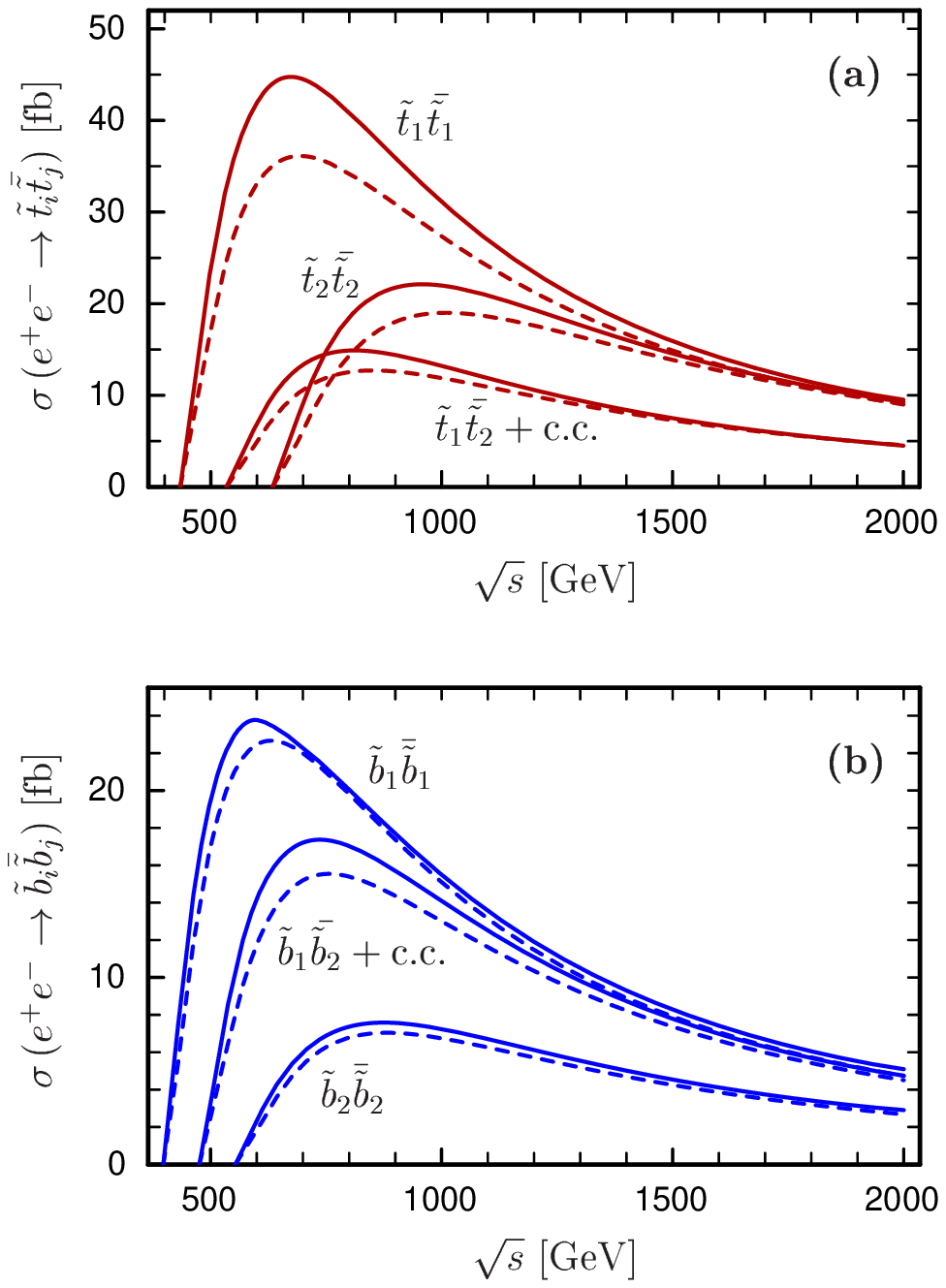}}}
\caption{Total (SUSY--QCD and Yukawa coupling) corrected cross sections 
(full lines) together with the tree--level cross sections (dashed lines) 
of (a) $e^+e^-\to\st_i\bar{\st_j}$ and (b) $e^+e^-\to\sb_i\bar{\sb}_j$
for the scenario of Table~\ref{tab:s1}.}
\label{fig:one}
}
 

\FIGURE{
	\centerline{\resizebox{13.5cm}{!}{\includegraphics{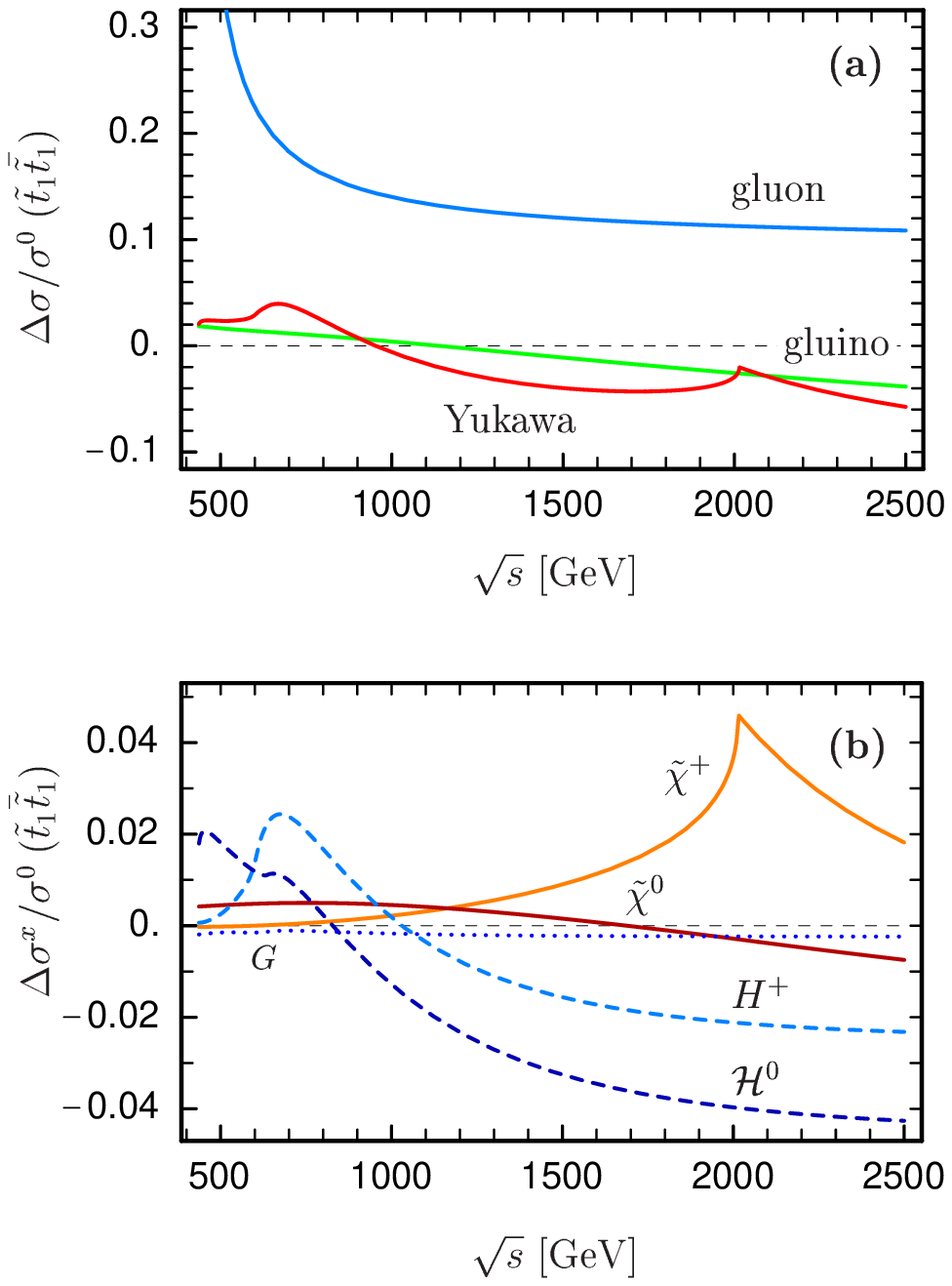}}}
\caption{Radiative corrections to $e^+e^-\to\st_1\bar{\st_1}$ 
relative to the tree--level cross section
for the scenario of Table~\ref{tab:s1}: 
(a) gluon, gluino, and Yukawa coupling corrections and 
(b) various Yukawa coupling correction contributions.}
\label{fig:two}
}


\FIGURE{
	\centerline{\resizebox{13.5cm}{!}{\includegraphics{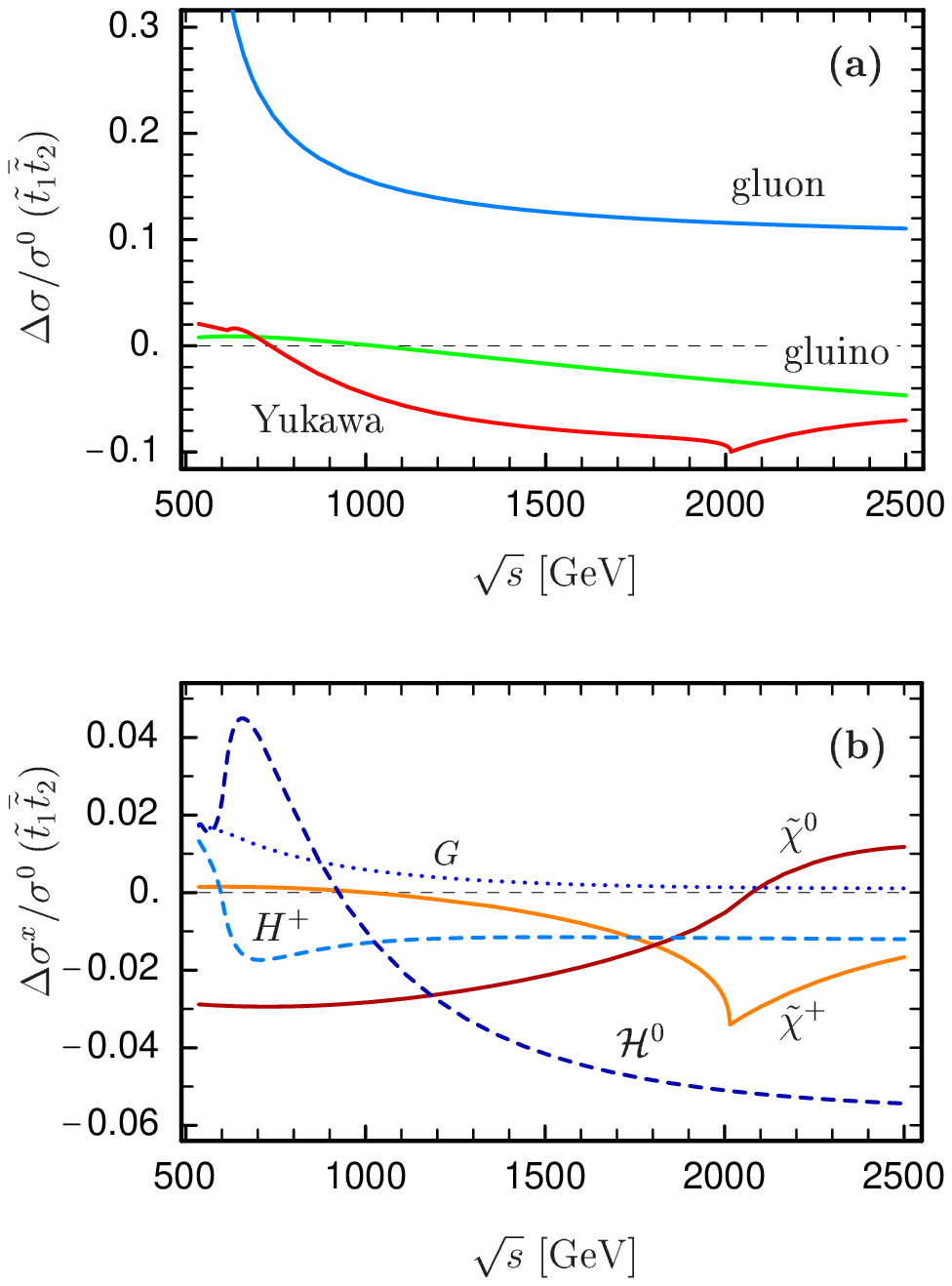}}}
\caption{Radiative corrections to $e^+e^-\to\st_1\bar{\st_2}+{\rm c.c.}$ 
relative to the tree--level cross section
for the scenario of Table~\ref{tab:s1}: 
(a) gluon, gluino, and Yukawa coupling corrections and 
(b) various Yukawa coupling correction contributions.}
\label{fig:three}
}


\FIGURE{
	\centerline{\resizebox{13.5cm}{!}{\includegraphics{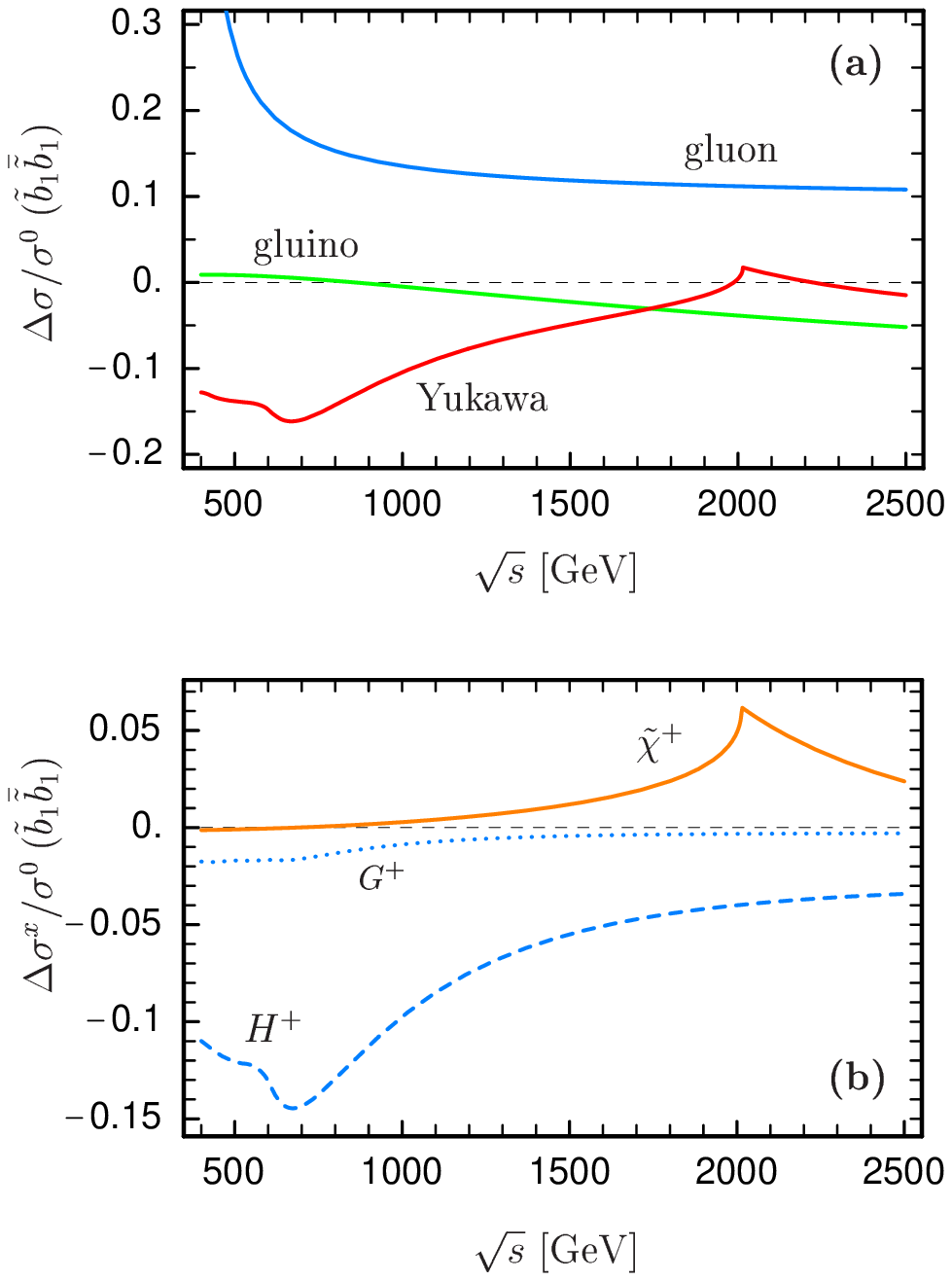}}}
\caption{Radiative corrections to $e^+e^-\to\sb_1\bar\sb_1$ 
relative to the tree--level cross section
for the scenario of Table~\ref{tab:s1}: 
(a) gluon, gluino, and Yukawa coupling corrections and 
(b) various Yukawa coupling correction contributions.}
\label{fig:four}
}


\FIGURE{
	\centerline{\resizebox{13.5cm}{!}{\includegraphics{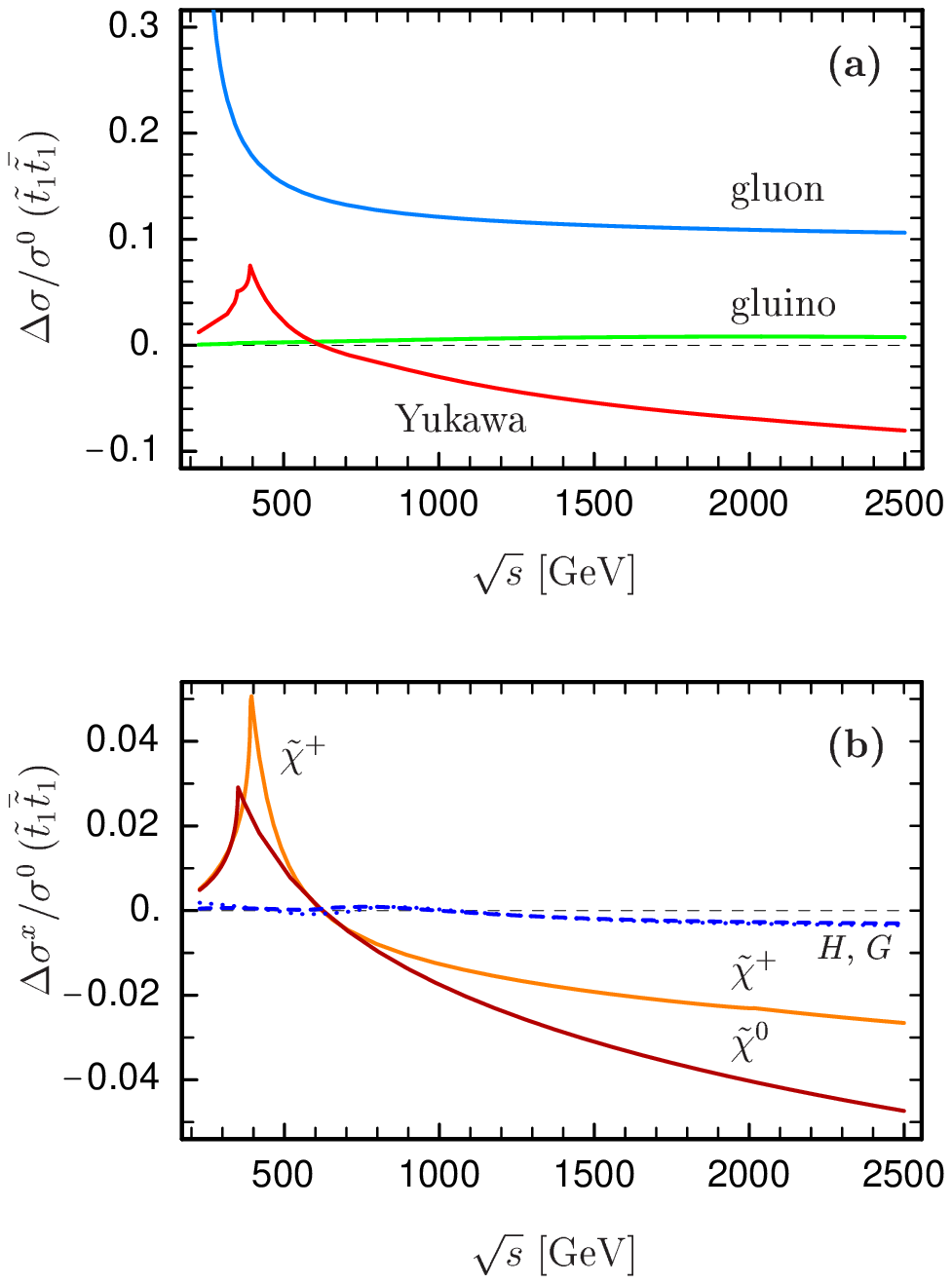}}}
\caption{Radiative corrections to $e^+e^-\to\st_1\bar{\st_1}$ 
relative to the tree--level cross section
for the scenario of Table~\ref{tab:s2}: 
(a) gluon, gluino, and Yukawa coupling corrections and 
(b) various Yukawa coupling correction contributions.}
\label{fig:five}
}


\FIGURE{
	\centerline{\resizebox{13.5cm}{!}{\includegraphics{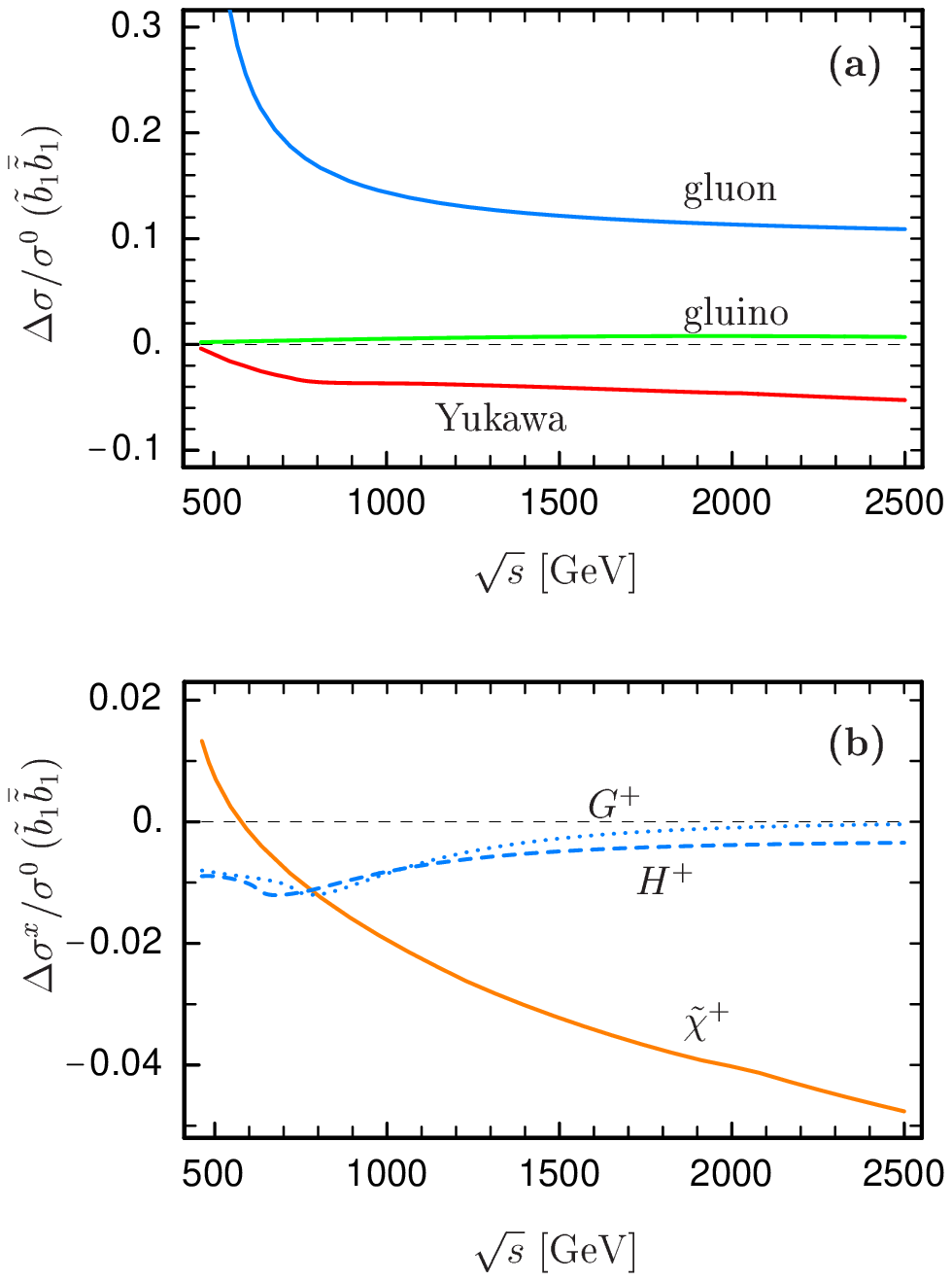}}}
\caption{Radiative corrections to $e^+e^-\to\sb_1\bar\sb_1$ 
relative to the tree--level cross section
for the scenario of Table~\ref{tab:s2}: 
(a) gluon, gluino, and Yukawa coupling corrections and 
(b) various Yukawa coupling correction contributions.}
\label{fig:six}
}


\FIGURE{
	\centerline{\resizebox{13.5cm}{!}{\includegraphics{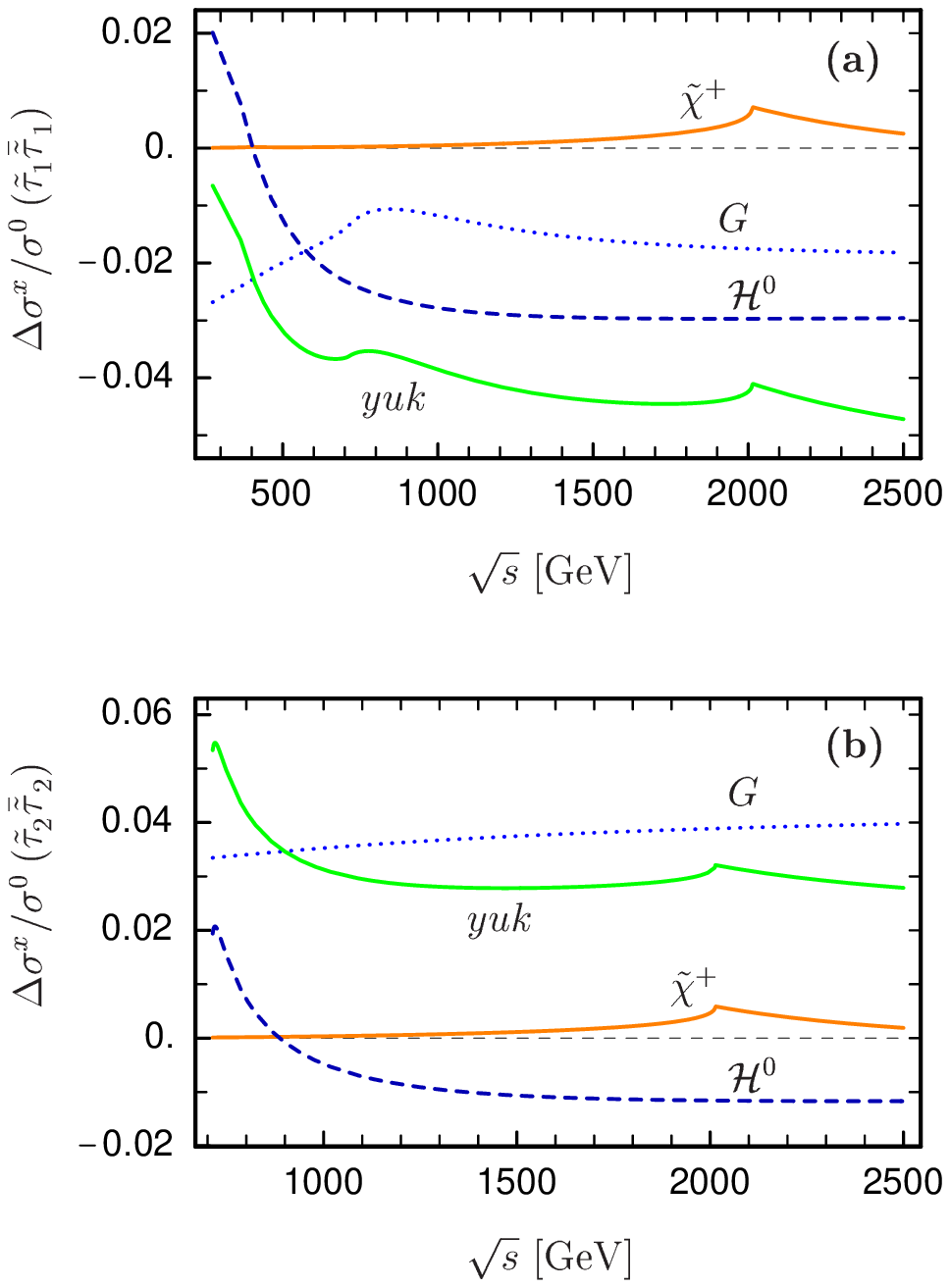}}}
\caption{Yukawa coupling correction contributions relative to
the tree--level cross section of 
(a) $e^+e^-\to\stau_1^{}\bar{\stau}_1^{}$ and 
(b) $e^+e^-\to\stau_2^{}\bar{\stau}_2^{}$ 
for $M_{\ti L}=280\gev$, $M_{\ti E}=250\gev$, $A_\tau=100\gev$,  
$M=200\gev$, $\mu=1000\gev$, $\tan\b=30$, $m_A=300\gev$.}
\label{fig:stau}
}


\begin{thebibliography}{99}
\bibitem{ref1}
H. P. Nilles, \prep{110}{1984}{1};\\ 
A. B. Lahanas and D. V. Nanopoulos, \prep{145}{1987}{1};\\ 
R. Barbieri, {\it Riv. Nuovo Cimento }{\bf 11} (1988) 1.

\bibitem{ref2}  
H. E. Haber and G. L. Kane, \prep{117}{1985}{75}.

\bibitem{ref3}   
J. F. Gunion, H. E. Haber, \npb{272}{1986}{1}.

\bibitem{ref4} 
A. Arhrib, M. Capdequi--Peyranere, A. Djouadi, \prd{52}{1995}{1404}.

\bibitem{ref4a}  
H. Eberl, A. Bartl, W. Majerotto, \npb{472}{1996}{481}.

\bibitem{ref5} 
W. Beenakker, R. H\"opker, and P. M. Zerwas, 
\plb{349}{1995}{463};\\
S.~Kraml, H.~Eberl, A.~Bartl, W.~Majerotto, and W.~Porod,\\
\plb{386}{1996}{175};\\
A. Djouadi, W. Hollik, and C. J\"unger, \prd{55}{1997}{6975}.

\bibitem{ref6}
W. Beenakker, R. H\"opker und P. M. Zerwas,
\plb{378}{1996}{159};\\
W. Beenakker, R. H\"opker, T. Plehn, and P.~M.~Zerwas,
\zpc{75}{1997}{349}.

\bibitem{ref7}
A.~Bartl, H.~Eberl, K.~Hidaka, S.~Kraml, W.~Majerotto, W.~Porod, and Y.~Yamada,
\plb{389}{1996}{538}.

\bibitem{ref8}
A. Arhrib, A. Djouadi, W. Hollik, and C. J\"unger,
\prd{57}{1998}{5860}.

\bibitem{ref9}
A.~Bartl, H.~Eberl, K.~Hidaka, S.~Kraml, W.~Majerotto, W.~Porod, and Y.~Yamada,
to be publ. in {\it Phys.~Rev.~D}. 

\bibitem{ref10}
A.~Bartl, H.~Eberl, K.~Hidaka, T.~Kon, W.~Majerotto, and Y.~Yamada, 
\plb{402}{1997}{303}.

\bibitem{ref11}
W.~Beenakker, R.~H\"opker, M.~Spira, and P.~M.~Zerwas,
\prl{74}{1995}{2905}, \zpc{69}{1995}{163},
\npb{492}{1997}{51}.

\bibitem{ref11a}
M.~A.~Diaz, S.~F.~King, and D.~A.~Ross, 
\npb{529}{1998}{23}.

\bibitem{ref11b}
S.~Kiyoura, M. M. Nojiri, D.~M.~Pierce, and Y.~Yamada,
\prd{58}{1998}{75002}.

\bibitem{ref12}
J. Guasch, J. Sol\`a, and W. Hollik
\plb{437}{1998}{88}.

\bibitem{ref13}
A. Akeroyd, A. Arhrib, and E. Naimi, hep-ph/9811431.

\bibitem{thesis}
H. Eberl, {\it Thesis}, University of Technology, Vienna, 1998.

\bibitem{tHooft}
G. 't~Hooft and M. Veltman, Nucl. Phys. B153 (1979) 365.

\bibitem{Denner}   
A. Denner, {\it Fortschr. Phys. }{\bf 41} (1993) 307.

\bibitem{charginos}
A. Bartl, H. Fraas, W. Majerotto, and B. M\"o\ss{}lacher,
\zpc{55}{1992}{257}.

\bibitem{neutralinos}
A. Bartl, H. Fraas, and W. Majerotto, 
\npb{278}{1986}{1}.

\bibitem{mh0}
J. Ellis, G. Ridolfi and F. Zwirner, \plb{262}{1991}{477}.

\bibitem{mhc}
A. Brignole, \plb{277}{1992}{313}.

\bibitem{lep2}
E.~Lancon (ALEPH), V.~Ruhlmann--Kleider (DELPHI), 
R.~Clare (L3) and D.~Plane (OPAL), 
talks at the 50th CERN LEPC meeting, 12 Nov. 1998; 
for minutes and transparencies, see 
{\tt http://www.cern.ch/Committees/LEPC/minutes/LEPC50.html}

\bibitem{tevatron}
D$\emptyset$ Collab., S. Abachi et al., 
\prl{76}{1996}{2222}; hep-ph/9902013; \\
CDF Collab., F. Abe et al., \prd{56}{1997}{1357}; \\
J. A. Valls, talk at the XXIX International Conference on High Energy Physics 
(ICHEP98), Vancouver, Canada, 23--29 July 1998, FERMILAB-Conf-98-292-E.

\bibitem{drhonum}
G. Altarelli, R. Barbieri and F. Caravaglios,
{\it Int. J. Mod. Phys.} {\bf A13} (1998) 1031.

\bibitem{Drees-Hagiwara}
M. Drees and K. Hagiwara, \prd{42}{1990}{1709}. 

\bibitem{Deren-Savoy}
J. P. Derendinger and C. A. Savoy, \npb{237}{1984}{307}.
 
\end{thebibliography}
\end{document}